### New Cataclysmic Variable 1RXS J161659.5+620014 in Draco

P. Balanutsa, D. Denisenko, E. Gorbovskoy, V. Lipunov

Sternberg Astronomical Institute at Lomonosov Moscow State University, Russia
e-mail: bala55@mail.ru, d.v.denisenko@gmail.com, gorbovskoy@gmail.com, lipunov2007@gmail.com

We report the discovery of a new cataclysmic variable MASTER OT J161700.81+620024.9 which is identical to the faint ROSAT X-ray source 1RXS J161659.5+620014. The object was observed in outbursts to $14.4^m$ on 2012 Sep. 11 by MASTER-Tunka and to $14.3^m$ on 2013 Jan. 21/22 by MASTER-Kislovodsk, but was not detected by the routine search procedures. Analysis of the archival MASTER data and CRTS light curve shows the variability from $17.8^m$ at quiescence to $14.3^m$ in outbursts, confirming that the new variable is a dwarf nova. SDSS colors suggest a small contribution from the secondary component and are telling in favor of the short orbital period.

New optical transients (OTs) discovered by MASTER global network of robotic telescopes (http://observ.pereplet.ru, Lipunov et al., 2010) are usually being published in the Astronomer's Telegram within a few hours to a few days after the detection by the automated search routines. However, some OTs (including the bright ones) were left unnoticed at the time of their outbursts due to various reasons described in Lazareva et al. (2013). Following the discovery of MASTER OT J015016.17+375620.5 = 1RXS J015017.0+375614 by A. Lazareva in MASTER-Amur data, we have conducted the search through MASTER-Kislovodsk and MASTER-Tunka databases for the missed OTs in the 30" circles around the faint ROSAT sources (Voges et al., 2000). As a result, we have found yet another bright ($14.3^m$) OT with two outbursts in 2012-2013.

The new object is named MASTER OT J161700.81+620024.9 according to its coordinates $16^h17^m00^s.81$, $+62°00'24''.9$ (J2000.0). It is located formally 14" from the X-ray source 1RXS J161659.5+620014 with the error radius of 9". The OT was detected in outburst by MASTER-Tunka on 2012 Sep. 11 with unfiltered magnitude $14.39^m$ and then again on 2013 Jan. 22 by MASTER-Kislovodsk at $14.43^m$. The detections by MASTER telescopes are listed in Table 1.

| Date, UT | Date, JD | Observatory | Mag | Error |
|---|---|---|---|---|
| 2012-09-11.536 | 2456182.036 | MASTER-Tunka | 14.39 | 0.01 |
| 2012-09-11.588 | 2456182.088 | MASTER-Tunka | 14.39 | 0.01 |
| 2013-01-21.987 | 2456314.487 | MASTER-Kislovodsk | 14.31 | 0.01 |
| 2013-01-22.017 | 2456314.517 | MASTER-Kislovodsk | 14.43 | 0.01 |
| 2013-01-27.707 | 2456320.207 | MASTER-Amur | 15.05 | 0.06 |
| 2013-01-27.739 | 2456320.239 | MASTER-Amur | 14.78 | 0.07 |

The comparison of MASTER-Tunka images taken on 2012 May 24 (with the object at quiescence) and on 2012 Sep. 11 (with the variable in outburst at $14.4^m$) is shown in Fig. 1.

The new variable is identical to the blue star USNO-B1.0 1520-0252429 with the coordinates $16^h17^m00^s.917$, $+62°00'24''.93$ (J2000.0), proper motion (14, 0) mas/yr in R.A. and Decl., respectively and the following magnitudes: $B1$=16.25, $R1$=16.56, $B2$=17.73, $R2$=16.80, $I$=16.35. Color-combined (BRIR) DSS finder chart is presented in Fig. 2 (10'x10' field of view). The star is listed in USNO-A2 catalogue as USNO-A2.0 1500-06057914 ($16^h17^m00^s.86$, $+62°00'24''.9$, $R$=16.6, $B$=16.2). There is also an ultraviolet counterpart in GALEX database GALEX J161700.9+620024 ($FUV$=17.82±0.06, $NUV$=17.76±0.04) and the infrared detection 2MASS 16170084+6200246 ($J$=16.37±0.13).

The X-ray source 1RXS J161659.5+620014 has a ROSAT flux 0.0179±0.0043 counts/s and hardness ratios HR1=0.52±0.25, HR2=0.60±0.20. It is remarkable that the X-ray flux of the new object is identical to that of 1RXS J015017.0+375614 (Lazareva et al., 2013). This value is typical for the dwarf novae of ~18$^{th}$ magnitude at quiescence and ~14$^m$ in outbursts. The hardness ratios also do not contradict to the dwarf nova classification of this object.

This area of sky in Draco was observed by Sloan Digital Sky Survey. The variable is present in release 9 of the SDSS photometric catalogue (Adelman-McCarthy et al., 2012) as SDSS J161700.91+620024.8 with the following magnitudes: $u$=16.43, $g$=16.14, $r$=16.21, $i$=16.35, $z$=16.41. Negative color indices ($g$-$r$) = –0.07, ($r$-$i$) = –0.14 suggest a small contribution from the secondary companion, telling in favor of a low mass red dwarf and short orbital period in this cataclysmic binary system.

The light curve of MASTER OT J161700.81+620024.9 = 1RXS J161659.5+620014 from the combined data of Amur, Tunka and Kislovodsk telescopes is shown in Fig. 3. Filled circles represent positive detections, triangles - upper limits.

We have also checked the available data on this object in Catalina Sky Survey database (Drake et al., 2009). The object was observed 93 times from 2006 May 01 to 2012 Sep. 27. One bright outburst was detected on 2010 June 19 (15.2$^m$), and another fainter one on 2012 Apr. 27 (15.6$^m$). The quiescent magnitude is 17.8$^m$. The light curve is typical for dwarf novae. Additional monitoring is required to detect the next outburst and to measure the orbital period which will allow making the final classification of this CV.

**Acknowledgements:** MASTER project is partially supported by State contract No. 11.G34.31.0076 and State Contract No. 14.518.11.7064 with Russian Ministry of Science and Education.

References:

Lipunov, V., Kornilov, V., Gorbovskoy, E., et al., 2010, *Advances in Astronomy*, article id 349171, p. 1
Lazareva, A., Voroshilov, N., Denisenko, D., et al., 2013, arXiv:1307.6855
Voges, W., Aschenbach, B., Boller, Th., et al., 2000, *IAU Circular* **7432**
Drake, A. J., Djorgovski, S. G., Mahabal, A., et al., 2009, *Astrophys. Journal*, **696**, 870
Adelman-McCarthy, J.K., et al., 2012, *Astrophys. Journal Suppl. Ser.,* **203**, 21

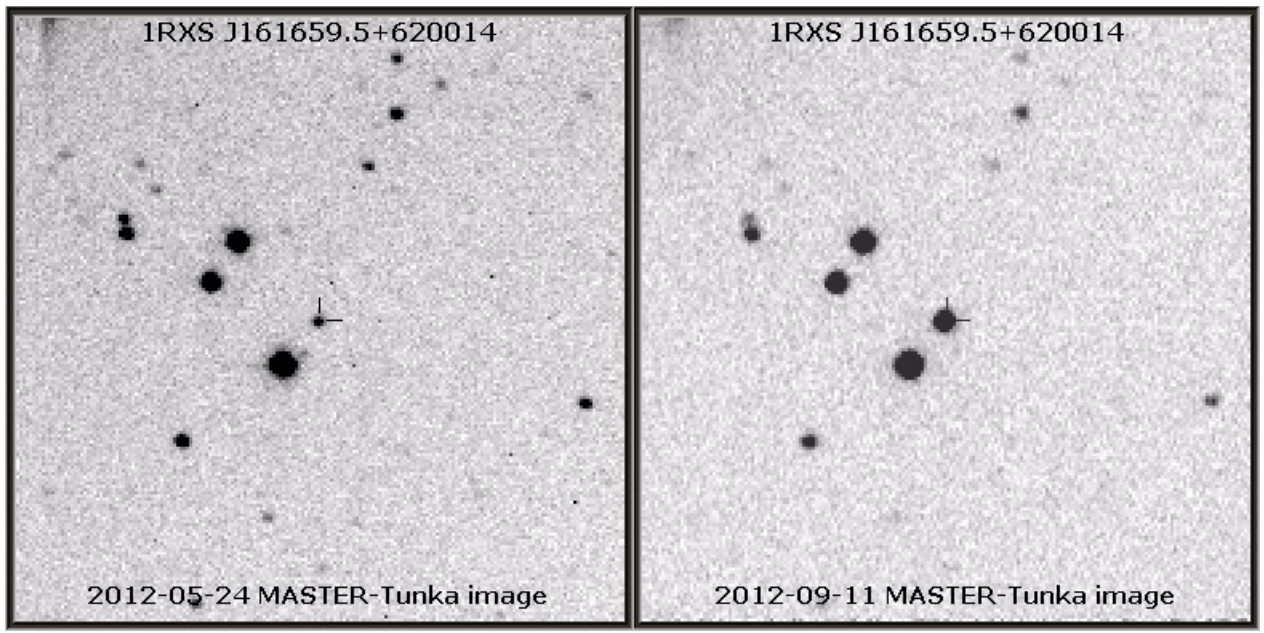

**Figure 1.** MASTER images of 1RXS J161659.5+620014 at quiescence on 2012 May 24 (left panel) and in outburst on 2012 Sep. 11 (right panel)

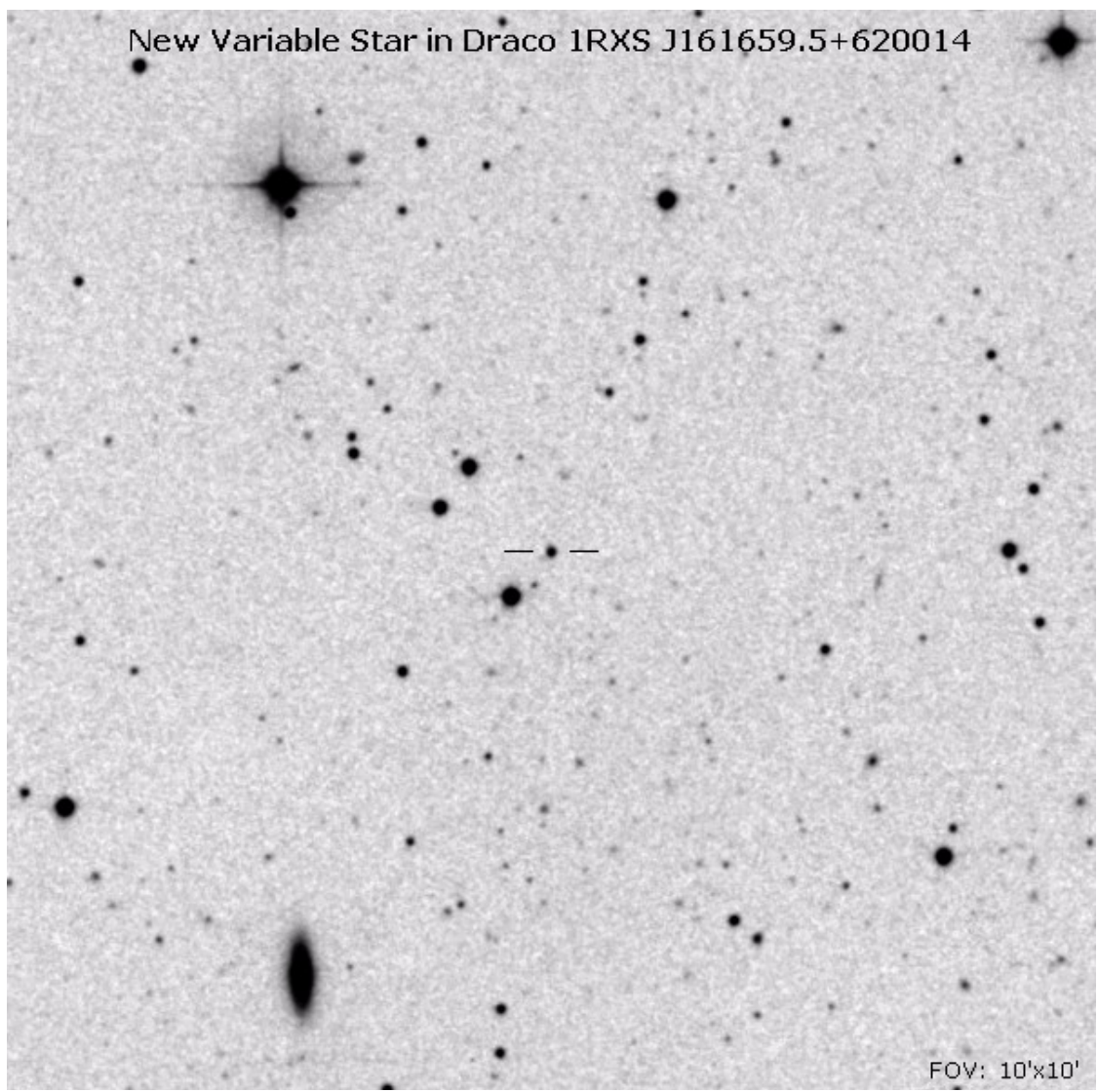

**Figure 2.** Combined DSS image of 1RXS J161659.5+620014 (10'x10' field of view)

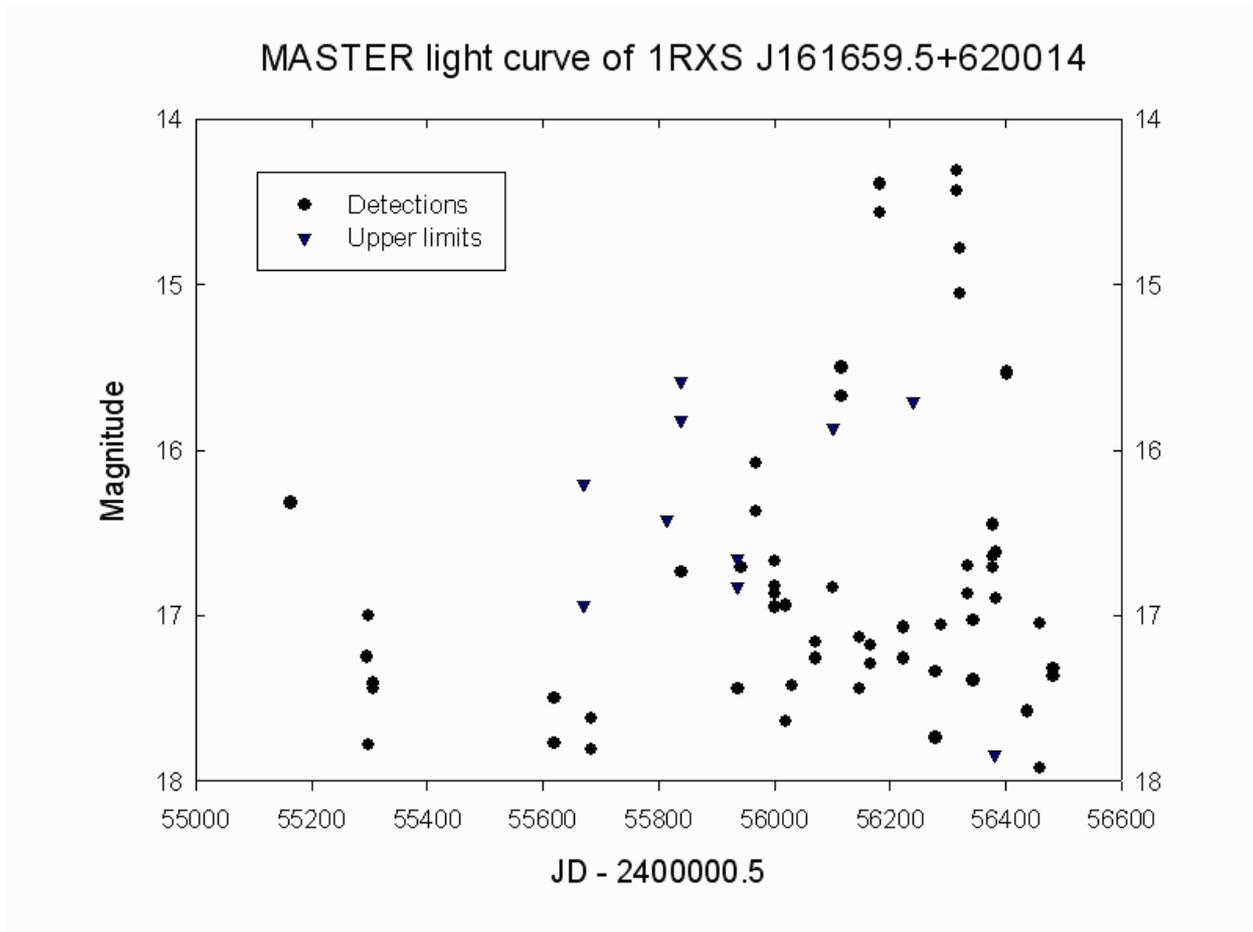

**Figure 3.** MASTER light curve of 1RXS J161659.5+620014. Circles are positive detection, triangles - upper limits.